\documentclass[twocolumn,aps,showpacs,prd]{revtex4}

\usepackage{amsmath}
\usepackage{amssymb}
\usepackage{amsfonts}
\usepackage{amsthm}
\usepackage{mathrsfs}

\newcommand{\bh}{\bar{h}}
\newcommand{\bT}{\bar{T}}

\newcommand{\hc}{\text{H.c.}}

\newcommand{\xx}{\boldsymbol{x}}
\newcommand{\kk}{\boldsymbol{k}}

\newcommand{\T}{\mathscr{T}}
\newcommand{\TT}{\text{TT}}
\newcommand{\dd}{\text{d}}
\newcommand{\CC}{{\mathcal C}}

\newcommand{\LLL}{\mathscr{L}}
\newcommand{\LL}{{\mathcal L}}

\newcommand{\HH}{{\mathcal H}}
\newcommand{\RR}{{\mathcal R}}
\newcommand{\DD}{{\mathcal D}}
\newcommand{\NN}{{\mathcal N}}
\newcommand{\UU}{{\mathcal U}}
\newcommand{\WW}{{\mathcal W}}

\newcommand{\bra}[1]{\langle#1|}

\newcommand{\ket}[1]{|#1\rangle}

\newcommand{\tr}{\text{tr}}
\newcommand{\ppp}{\\}
\newcommand{\nppp}{\nonumber\\}

\begin{document}

\title{Quantum gravitational decoherence of light and matter}

\author{Teodora Oniga}
\email{t.oniga@abdn.ac.uk}

\author{Charles H.-T. Wang}
\email{c.wang@abdn.ac.uk}
\affiliation{Department of Physics,
University of Aberdeen, King's College, Aberdeen AB24 3UE, United Kingdom}


\begin{abstract}
Real world quantum systems are open to perpetual influence from the wider environment. Quantum gravitational fluctuations provide a most fundamental source of the environmental influence through their universal interactions with all forms of energy and matter causing decoherence. This may have subtle implications on precision laboratory experiments and astronomical observations and could limit the ultimate capacities for quantum technologies prone to decoherence. To establish the essential physical mechanism of decoherence under weak spacetime fluctuations, we carry out a sequence of analytical steps utilizing the Dirac constraint quantization and gauge invariant influence functional techniques, resulting in a general master equation of a compact form, that describes an open quantum gravitational system with arbitrary bosonic fields. An initial application of the theory is illustrated by the implied quantum gravitational dissipation of light as well as (non)relativistic massive or massless scalar particles. Related effects could eventually lead to important physical consequences including those on a cosmological scale and for a large number of correlated particles.
\end{abstract}


\maketitle

\section{Introduction}

Theories of gravitational decoherence have become increasingly important in the study of quantum systems, as gravitation can never be turned off. The ever more accurate quantum measurements may therefore be holding out the prospects of probing new physics normally attached to Planck scale physics and quantum cosmology \cite{lammerzahl, kiefer2}.
Decoherence plays a central role in quantum-to-classical transition, a main question in the foundation of quantum mechanics, through unavoidable entanglements between the quantum system and the environment, leading to the loss of coherence between the system's superposition states and hence the appearance of a classical mixture \cite{Giulini, Schlosshauer}.
Gravitationally induced decoherence could therefore set a limit for precision measurements and astronomical observations providing a strong motivation to investigate its full nature and detailed mechanisms \cite{lamine, wang, bonifacio, amelino, schiller, pikovski2}.

The issue of decoherence due to gravity has been investigated  in terms of two different classes of models. First, intrinsic or fundamental decoherence \cite{anastopouloshu3} suggests a new process intrinsic to all quantum systems, whereby decoherence occurs through spontaneous wavefunction collapse, typically involving modifications of Schr\"{o}dinger's equation \cite{ghirardi2, ghirardi3}. In particular, it has been proposed that gravity can be the cause of this wavefunction collapse \cite{karolyhazy, ghirardi, penrose, diosi, bassi, adler}.
Second, environmentally induced decoherence refers to the influence of the environment on a quantum system through their interaction, leading to the destruction of superposition states. There have been a number of approaches to environmental gravitational decoherence in the recent literature, with various sources of decoherence, including semiclassical metric fluctuations \cite{wang, bonifacio, breuer1, ford1}, quantized Newtonian gravity \cite{kay}, gravitational time dilation \cite{pikovski1} and thermal spacetime foam \cite{garay}. Decoherence due to low-energy gravity has also been studied in \cite{anastopouloshu, blencowe, anastopoulos}.
While these two classes of decoherence are fundamentally different, it is interesting to seek justifications for some of the prescriptions of intrinsic decoherence from gravitational decoherence on phenomenological grounds \cite{power}.

Quantum fluctuation phenomena such as the Lamb shift, Casimir effect and spontaneous emission have been crucial in the development and interpretation of QED and the quantum theory as a whole \cite{Giulini, Schlosshauer, Boi}. Decoherence due to electromagnetic fluctuations as a potential new quantum vacuum effect has long attracted attention with pioneering analysis reported in \cite{ford1993}. Theoretically, spacetime fluctuations must also exist due to the quantization of the metric field \cite{ford2, ford3}, though details may vary with presently uncertain theories of quantum gravity. In the linearized approximation of gravity, one hopes to draw valuable analogy with QED despite differences in certain details. For example, gravitational vacuum could not be easily modified for testing as with the electromagnetic vacuum using small plates suggested in \cite{ford1993}.

Metric fluctuations can be passively induced by the quantum fluctuations of matter fields, with a significant amount of work already done \cite{sgrav}.
Treating spacetime itself as a noisy environment involves active metric fluctuations originating from the quantum nature of the true gravitational degrees of freedom \cite{wang2}, which we will investigate through gravitational decoherence. Here we focus on the active quantum environmental metric fluctuations without using semiclassical stochastic approximations.

\newpage

Through previous works, substantial increase of understanding and demonstration of a broad range of possible important and novel gravitational effects have been achieved. To advance forward, however, a conceptually clear and mathematically tractable theoretical description is in order, providing at one's disposal a framework and tools readily adapted to the analysis of gravitational decoherence in such fields as quantum computing, dynamics, information, gravity, metrology, and optics. The required theory should be able to model known forms of matter and arbitrary number of particles yet free from particular reference frames, so that such restrictions to varied degrees experienced by previous approaches may be circumvented. For expected weak gravitational decoherence, these requirements could be achieved using a Lorentz and gauge invariant quantum field-theoretical description.

In this paper we set out to construct {\it ab initio} a generic gravitational master equation with minimal assumptions.
Environmental gravitational decoherence preserving gauge invariance is addressed using Dirac quantization \cite{Dirac1964} and an influence functional technique in the framework of open quantum systems
\cite{oqs, breuer1, pikovski1, anastopouloshu}.
The resulting master equation is reported for general bosonic fields. We conclude with remarks and interpretations of this master equation and illustrate its application in the quantum dissipations of photons and scalar particles, in relativistic and nonrelativistic regimes.

Unless otherwise noted, we adopt units with unity speed of light $c=1$ preserving other constants including Newton's $G$, Planck's $\hbar$ and Boltzmann's $k_B$.
Commutators, anti-commutators, and Poisson brackets are denoted by
$[\cdot, \cdot ], \{\cdot, \cdot \}$ and $\{\cdot, \cdot \}_P $ respectively.
Spacetime and spatial coordinates are labelled by Greek and Latin letters starting from 0 (for time) and 1 respectively. Summing over repeated coordinate and polarization indices is implied. Spacetime coordinates $(x^\mu)$ are sometimes denoted by $(\xx,t)$ or simply $(x)$. In terms of the background Minkowski metric $\eta_{\mu\nu} = $diag$(-1,1,1,1)$, used to shift indices, and metric perturbation $h_{\mu\nu}$, the spacetime metric follows from standard linearized gravity notation as $g_{\mu\nu} = \eta_{\mu\nu} + h_{\mu\nu}$.

\section{Quantization of linearized gravity with matter}

Let us start from the total Lorentz invariant Lagrangian density $\LL =\LL_g + \LL_\varphi$ in terms of $\LL_g={\RR}/\kappa$ with $\kappa={16\pi G}$ and quadratic scalar curvature $\RR$ for linearized gravity and $\LL_\varphi$ depending on a (multiple-component) bosonic field $\varphi$ and up to linear terms in the metric perturbation $h_{\mu\nu}$.
Denoting differentiation with a comma, the metric perturbation undergoes the gauge transformation
$
h_{\mu\nu} \rightarrow h_{\mu\nu} + \xi_{\mu,\nu} + \xi_{\nu,\mu}
$
derived from the coordinate transformation
$x^\mu\to x^\mu-\xi^\mu$ with any small $\xi^\mu=(\xi,\xi^i)$.

To formulate a gauge and hence coordinate independent description of quantum gravitational decoherence, we invoke Dirac's theory \cite{Dirac1964} of constrained Hamiltonian systems and their quantization.
The gauge invariant description of free gravitons in vacuum using the particle representation has been obtained in \cite{kuchar}. Here gauge invariant interactions with matter fields will be further incorporated so as to account for their gravitational decoherence. By virtue of a set of primary constraints, the canonical variables of the metric are given by $h_{ij}$ and their conjugate momenta $p_{ij}=\partial\LL_g/\partial \dot{h}_{ij}$, with an over-dot denoting time derivative.
The remaining $h_{0\mu}$ become Lagrangian multipliers
$n=-h_{00}/2$ and $n_i = h_{0i}$. The corresponding first class Hamiltonian and momentum constraints are given respectively by
$
\CC_g
=
( h_{ii,jj} - h_{ij,ij})/\kappa
,\,
\CC_g^i
=
-2p_{ij}{}_{,j}
$.
They enter into the extended Hamiltonian for linearized gravity $H_g = \int \HH_g \,\dd^3x$ with density $\HH_g = p_{ij}\dot{h}_{ij} - \LL_g$.
The linearized Einstein tensor $G_{\mu\nu}$ can be expressed as
\begin{eqnarray}
&&
G_{00}
=
-\frac{\kappa}2\,\CC_g
,\;
G_{0i}
=
-\frac{\kappa}2\,\CC_g^i
\nppp&&
G_{ij}
=
\kappa \big[ \dot{p}_{ij} - \{ p_{ij}, H_g \}_P \big].
\label{eintenija}
\end{eqnarray}

In canonical general relativity, the Hamiltonian $H_\varphi=\int \HH_\varphi \,\dd^3x$ of the field $\varphi$ with conjugate momentum $\varpi$ has the density of the ADM form \cite{adm}
$
\HH_\varphi
=
N \CC_\varphi(\varphi,\varpi,g_{kl}) + N_i\, \CC_\varphi^i(\varphi,\varpi,g_{kl})
$
where $N$ is the lapse function, $N_i$ is the shift vector with functions
$\CC_\varphi$ and $\CC_\varphi^i$ independent of the derivatives of $g_{kl}$ as the matter part of Hamiltonian and momentum constraints respectively.

To proceed, $\HH_\varphi$ is linearized by using
$N=1 + n$, $N_i = n_i$ and $g_{ij} = \eta_{ij} + h_{ij}$
and keeping up to linear order of $n, n_i$ and $h_{ij}$. This yields
$
\HH_\varphi
=
\HH_S + \HH_I
$
where
$\HH_S = \CC_\varphi|_{h = 0}$
gives the Minkowski Hamiltonian
$H_S=\int\HH_S\,\dd^3 x$
describing the reduced system and the interaction Hamiltonian density
\begin{eqnarray}
\HH_I = -\frac12\, h_{\mu\nu}   T^{\mu\nu}
\label{HHI}
\end{eqnarray}
for gravitational couplings of this system, involving the Minkowski stress-energy tensor
\begin{eqnarray}
&&
T^{00} = \CC_\varphi|_{h = 0}
\,,\;
T^{0i} = -\CC_\varphi^i|_{h = 0}
\nppp&&
T^{ij}
=
-2\left.{\partial\CC_\varphi}/{\partial g_{ij}}
\right|_{h = 0} \,.
\label{CTij}
\end{eqnarray}
Importantly, the gauge invariance of the interaction Hamiltonian $H_I=\int\HH_I\,\dd^3 x$
follows from the conservation law
$\partial_\mu  T^{\mu\nu} = 0$ also required by the linearized Einstein equation.

Subsequently, gauge transformations of canonical variables of the gravity-matter system are equivalent to canonical transformations generated by the smeared constraint
$C = \int(\xi\CC + \xi_i\CC^i)\,\dd^3 x$
using the total Hamiltonian and momentum constraints
\begin{eqnarray}
\CC=\CC_g + \CC_\varphi
\,,\;
\CC^i=\CC_g^i + \CC_\varphi^i
\label{GMCC}
\end{eqnarray}
entering the total Hamiltonian $H = H_g+H_\varphi$.
Using \eqref{eintenija} and \eqref{CTij} we see that, respectively, the $(00)$ and $(0i)$-components of the linearized Einstein equation
\begin{eqnarray}
G_{\mu\nu} = \frac\kappa{2}\,  T_{\mu\nu}
\label{eineqb}
\end{eqnarray}
are equivalent to $\CC=0$ and $\CC^i=0$ with the remaining satisfied by the canonical field equations for $h_{ij}$ and $p_{ij}$.

On Dirac quantization in the Heisenberg picture, fields $h_{ij}, \varphi$ and their momenta become operators satisfying the corresponding canonical commutation relations. They evolve according to the Heisenberg equations:
\begin{eqnarray}
&&
\dot{h}_{ij}
=
-\frac{i}{\hbar} [h_{ij}, H]
\,,\;
\dot{p}_{ij}
=
-\frac{i}{\hbar} [p_{ij}, H]
\label{pHH}
\ppp&&
\dot\varphi = -\frac{i}{\hbar} [\varphi, H]
\,,\;
\dot\varpi = -\frac{i}{\hbar} [\varpi, H]
\label{cpiHH}
\end{eqnarray}
using the operator form of $H$.
The operator forms of constraints $\CC$ and $\CC^i$ with consistent factor ordering become quantum generators of gauge transformation requiring physical states $\ket{\psi}$ to be gauge invariant by satisfying
\begin{eqnarray}
\CC \ket{\psi}=0
\,,\;
\CC^i \ket{\psi}=0 \,.
\label{qCCipsi}
\end{eqnarray}

The equations in \eqref{pHH} can be shown to be fully equivalent to the $(ij)$ components of  \eqref{eineqb} as quantized operator equations. Moreover, let us consider only gauge invariant states $\ket{\psi}$ satisfying \eqref{qCCipsi} and use them to evaluate matrix elements of operators. In this sense, the quantum evolution of the total system is determined by the matter Heisenberg equations \eqref{cpiHH} and the quantized Einstein equation \eqref{eineqb}.

Because of gauge redundances, supplementary relations, which may be called quantum gauge conditions, can be applied to field operators to isolate dynamical degrees of freedom. Different from the usual gauge fixing by setting constraints to zero classically already, however, we reserve the freedom to change gauge that can be generated by the constraint operators. To this end, we apply the operator Lorenz condition $\bh_{\mu\nu}{}^{,\nu}=0$, with an over-bar denoting trace-reversion. The quantum Einstein equation then reduces from the form of \eqref{eineqb} to
\begin{eqnarray}
h_{\mu\nu,\alpha}{}^\alpha = - \kappa  \bT_{\mu\nu} \,.
\label{lfeq3}
\end{eqnarray}
The solution to the above naturally separates into
$
h_{\mu\nu} = u_{\mu\nu} + \gamma_{\mu\nu}
$
in terms of the retarded potential
\begin{eqnarray}
u_{\mu\nu}(\xx,t) =
\frac\kappa{4\pi}\int \dd^3 x'
\frac{\bT_{\mu\nu}(\xx',t-|\xx-\xx'|)}{|\xx-\xx'|}
\label{ret}
\end{eqnarray}
augmented by any boundary terms,
and $\gamma_{\mu\nu}$ satisfying the homogeneous part of \eqref{lfeq3}.
It is then cast into a transverse-traceless (TT) form
$\gamma_{\mu\nu}=\gamma_{\mu\nu}^\text{TT}$
with a Lorenz-preserving coordinate transformation.
Clearly $\gamma_{ij}$  carries the dynamical degrees of freedom of gravity satisfying the operator wave equation
$\gamma_{ij,00}-\gamma_{ij,kk} = 0$.

Using \eqref{ret}, the orthogonality of TT decomposition \cite{York1974} and the densities
\begin{eqnarray}
\UU
=
-\frac{\kappa}{8\pi}\!\int\!\dd^3 x'
\frac{T^{\mu\nu}(\xx,t)\bT_{\mu\nu}(\xx',t-|\xx-\xx'|)}
{|\xx-\xx'|}
\label{HIU}
\end{eqnarray}
and
\begin{eqnarray}
\WW = -\frac12\gamma_{ij}\tau_{ij}
\label{HIW}
\end{eqnarray}
we can split  $H_I = U + W$, where $U = \int\UU\,\dd^3 x$ arises from self-gravity and $W = \int\WW\,\dd^3 x$ in terms of the TT stress tensor $\tau_{ij}=T^\TT_{ij}$ which describes the actual environmental coupling with the  gravitational wave field $\gamma_{ij}$.
The operator wave equation for $\gamma_{ij}$ is solved by
\begin{eqnarray}
\gamma_{ij}(x)
=
\int\!\dd^3 k
\sqrt{\frac{\kappa\hbar}{2(2\pi)^3 k}}\,
g_{\kk}^\lambda\,
e_{ij}^\lambda(\kk)\,
e^{ikx}
+\hc
\label{wex}
\end{eqnarray}
where $kx = \kk\cdot \xx - \omega t$ with $\omega=k=|\kk|$,
$e_{ij}^\lambda(\kk)$ are basis TT tensors relative to $\kk$ with two polarizations $\lambda=1,2$, and $\hc$ signifies Hermitian conjugate. They are normalized with $e_{ij}^\lambda(\kk)e_{kl}^\lambda(\kk)=2\,P_{ijkl}(\kk)$ using
the TT projector
$
P_{ijkl}
=
(P_{ik}P_{jl} + P_{il}P_{jk} -  P_{ij}P_{kl})/2
$
in terms of the transverse projector $P_{ij}$.
The canonical commutation relations for gravity then require
$
\big[
g_{\kk}^\lambda, g_{\kk'}^{\lambda'\dag}
\big]
=
\delta_{\lambda\lambda'}\delta^3(\kk, \kk')
,\,
\big[
g_{\kk}^{\lambda\dag}, g_{\kk'}^{\lambda'\dag}
\big]
=
\big[
g_{\kk}^\lambda, g_{\kk'}^{\lambda'}
\big]
=0
$, which after some calculations allows the normal-ordered gravitational Hamiltonian to take the canonical form:
\begin{eqnarray}
H_g
=
\int\!\dd^3 k\,
\hbar\,k\,
g_{\kk}^{\lambda\dag}g_{\kk}^{\lambda} \,.
\label{gham}
\end{eqnarray}
Consequently, the properties of gravitons are analogous to photons, with $g_{\kk}^{\lambda\dag}$ and $g_{\kk}^{\lambda}$ as creation and annihilation operators for gravitons. The gauge invariance of graviton states such as
$\ket{\psi} = g_{\kk}^{\lambda\dag} \ket{0}$ follows from that of the TT metric perturbation implying $\gamma_{ij}$ commutes with $\CC$ and $\CC^i$ thereby \eqref{qCCipsi} is satisfied.

\vspace{40pt}

\section{Derivation of the general gravitational master equation}

Interactions with the gravitational environment will lead to nonunitary dynamics of the reduced matter system that can be treated as an open quantum system, described by a total Hamiltonian of the form $H = H_S + H_I + H_B$. Such an open quantum gravitational system emerges unambiguously from the present matter-gravity Hamiltonian $H = H_\varphi + H_g$ through $H_\varphi = H_S + H_I$ derived earlier and now completed with the bath Hamiltonian $H_B = H_g$ specified in \eqref{gham}. That $H$ is gauge invariant follows from the gauge invariant constructions of $H_S$, $H_I$ and $H_B$. By tracing out the environment degrees of freedom from the total density matrix $\varrho$ we obtain the equation of motion for the reduced density matrix $\rho$. For this purpose, it is useful to switch over to the interaction picture where $H_I$ generates the time evolution of quantum states.
We then turn to the Liouville-von Neumann equation describing the total evolution
\begin{equation}
\dot\varrho(t)
=
\int\!\dd^3 x\LLL_{\HH_I}(x) \varrho(t)
\label{drho}
\end{equation}
using the Liouville super-operator notation so that
$\LLL_{A}B = - (i/ \hbar) [A,B]$ for operators $A$ and $B$.

At an initial time conveniently set to $t=0$, the total system takes a separable form $\varrho(0) = \rho(0) \otimes \rho_B$ where $\rho_B$ describes the gravitational thermal bath assumed to be unaffected by its weak coupling to matter, though matter dynamically develops entanglements with gravity. Based on \eqref{HIU} and \eqref{HIW}, the iterative solution of \eqref{drho}, after applying a partial trace $\tr_B$ over the bath, is given by
\begin{eqnarray}
&&\!\!\!\!\!\!\!\!
\rho(t) =
\T_\tau \bigg\{ \exp \Big [ \int_{0}^{t} \dd^4 x' \LLL_\UU(x') \Big ] \times
\nppp
&&
\tr_B \Big \{ \T_\gamma \exp \Big [ \int_{0}^{t} \dd^4 x' \LLL_\WW(x')  \Big ] \varrho(0) \Big \} \bigg\}
\label{rhoeq}
\end{eqnarray}
using time ordering $\T=\T_\gamma \T_\tau$ for the graviton field $\gamma_{ij}$ and TT stress tensor $\tau_{ij}$ respectively, as the two sets commute.
The second factor of \eqref{rhoeq} is evaluated using
\begin{eqnarray}
&&\!\!\!\!\!\!\!\!\!
\T_\gamma \exp \Big [ \int_{0}^{t} \dd^4 x' \LLL_\WW(x') \Big ]=
\nppp&&
\exp \Big [ \frac {1}{2} \int_{0}^{t}\! \dd^4 x'\! \int_{0}^{t}\! \dd^4 x''
[\LLL_\WW(x'), \LLL_\WW(x'')] H(t'-t'') \Big ]
\nppp&&
\times \,\exp \Big [ \!\int_{0}^{t}\! \dd^4 x \LLL_\WW(x') \Big ]
\label{Texp}
\end{eqnarray}
with the Heaviside function $H(t)$.

In analogy with the QED theory with a linear field dependence of the interaction Hamiltonian, here the gravitational bath state is also assumed to be Gaussian  \cite{ferraro}.
In particular, it is physically reasonable to assume the gravitational environment to be in thermal equilibrium having a Planck distribution
$N(\omega)=1/(e^{\hbar\omega/k_B T}-1)$ of gravitons with frequency $\omega$ at temperature $T$.

Using the ensemble average over bath $\langle\cdot\rangle_B$,
the influence functional can then be written as a cumulant expansion \cite{oqs} for \eqref{Texp} which vanishes after the second order \cite{cumulant} such that
\begin{eqnarray}
&&\!\!\!\!\!\!\!
\tr_B\left \{ \exp \Big [ \int_{0}^{t} \dd^4 x' \LLL_\WW (x') \Big ] \rho(0) \right \} =
\nppp
&& \exp \left\{ \frac{1}{2} \int_{0}^{t} \dd^4 x' \int_{0}^{t} \dd^4 x'' \langle \LLL_\WW (x') \LLL_\WW (x'')\rangle_B \right\} \rho(0) .
\nppp&&
\label{trb}
\end{eqnarray}
Substituting the expansion of \eqref{trb} into \eqref{rhoeq}, and using the dissipation $\DD(x)$ and noise $\NN(x)$ kernels satisfying
\begin{eqnarray*}
[\gamma_{ij}(x),\gamma_{kl}(x')]
&=&
2 i \kappa\hbar \, P_{ijkl} \DD(x-x')
\ppp
\langle  \{ \gamma_{ij}(x) , \gamma_{kl}(x') \} \rangle_B
&=&
2 \kappa\hbar  P_{ijkl} \NN(x-x')
\end{eqnarray*}
we see that, after a lengthy calculation, the reduced density matrix takes the form $\rho(t)= \T_\tau e^{i \Phi} \rho(0)$, using the gravitational influence phase functional $\Phi$ so that:
\begin{eqnarray}
&&\!\!\!\!\!\!\!\!\!\!\!
i\Phi\rho
=
\int_{0}^{t} \dd^4 x \LLL_\UU(x)\rho
-\frac{\kappa}{4 \hbar} \int_{0}^{t} \dd^4 x' \int_{0}^{t} \dd^4 x''
\nppp
&&\;\;\;
\Big \{ i \DD(x'-x'') \Big[\tau_{ij}(x'),\big\{\tau_{ij}(x''),\rho\big\}\Big]
\nppp
&&\;\;\;
+  \NN(x'-x'')  \Big[\tau_{ij}(x'),\big[\tau_{ij}(x''),\rho\big]\Big] \vphantom{\int_{0}^{t_f}} \Big \} .
\label{Phi}
\end{eqnarray}
Using the integrals
$
\tau_{ij}(\kk, t)
=
\int\dd^3x \,
\tau_{ij}(\xx, t)
e^{-i \kk\cdot\xx}
$
and
$
\tilde{\tau}_{ij}(\kk, t)
=
\int_{0}^{t} \dd t'\,
\tau_{ij}(\kk, t')
e^{-i k (t - t')}
$,
we obtain with some more algebra the following exact nonlocal relation that
the time derivative of the above $\rho(t)$ satisfies:
\begin{eqnarray}
&&\!\!\!\!\!\!\!\!\!\!\!\!\!\!
\dot\rho(t)
=
-\frac{i}{\hbar} [U, \rho(t)]
\nppp&&
-
\frac{8\pi G}{\hbar}
\int\! \frac{\dd^3 k}{2(2\pi)^3k}
\Big \{
\big[
\tau^\dag_{ij} (\kk, t),\,
\tilde{\tau}_{ij}(\kk, t) \rho(t)
\big]
\nppp
&&
+N(k)
\big[
\tau^\dag_{ij} (\kk, t),\,
\left[
\tilde{\tau}_{ij}(\kk, t),\,
\rho(t)
\right]\big]
+\hc
\Big\}
\label{maseqn}
\end{eqnarray}
which is the sought master equation for a matter system subject to weak spacetime fluctuations described by linearized general relativity.

The consistency with gravitational gauge invariance is ensured by the use of gauge invariant operators and states in \eqref{maseqn}. The first term of this equation is due to the passive self-gravity of matter and retains unitary quantum nature. The second term of \eqref{maseqn} has nonunitary characteristics and therefore can induce decoherence and dissipation. In turn, this dissipator consists of the vacuum contribution independent of temperature due to zero-point spacetime fluctuations. The other term, proportional to the Planck distribution $N(k)$ of environmental gravitons, is the thermal contribution to quantum gravitational decoherence and dissipation.

The present formalism is applicable for general bosonic fields since their gravitational interactions do not involve derivatives of the metric. An extended description to accommodate fermions without this simplifying restriction is a subject for future work. As with standard open quantum systems \cite{oqs}, the reduced matter system is understood in the sense of effective quantum field theory up to a UV cutoff scale $\Omega$ due to its physical preparation and phenomenological constraints \cite{oqs}. This cutoff need not violate Lorentz invariance as it is defined relative to the centre of mass of the system. Such a system is said to be in a dressed state \cite{referee} with particle modes above the $\Omega$ scale contributing to nondissipative renormalization of physical parameters through vacuum polarizations. Therefore, the vacuum contribution of the master equation \eqref{maseqn} is subject to cutoff $\Omega$.

While some authors argue that $\Omega$ may well turn out to be zero, diminishing completely decoherence due to vacuum fluctuations  \cite{kiefer}, other authors consider the cutoff value to depend on the specific environment \cite{zeh}. For gravitational decoherence, without the availability of full quantum gravity theory, we tentatively adopt a phenomenological approach that $\Omega$ should not exceed the available energy scale of the reduced matter system. This is consistent with the Compton cutoff for nonrelativitic particles \cite{breuer1, oqs, anastopoulos}. Furthermore, for relativistic and massless particles, the cutoff should not exceed the maximum energy of their source.

For a bound matter system, additional approximations may be applied to \eqref{maseqn}, e.g. Born, Markov and rotating wave approximations in the optical limits, given justified time scales. We defer these investigations though they could bear significant physical consequences. For an unbound system such as free particles, the master equation \eqref{maseqn} describes their quantum Brownian motion and the resulting decoherence and dissipation due to spacetime fluctuations.

\section{Discussions}

Formulating a gravitational master equation has been a subject of some substantial investigations. See e.g. \cite{anastopouloshu, hu2, anastopoulos} and many related references therein. For environmentally induced decoherence,
simple forms of matter such as scalar fields or point masses coupled to gauge fixed gravitational fields, with applications in the nonrelativistic regimes, have been commonly used in these previous works.
In contrast, Eq.~\eqref{maseqn} is free from these restrictions. The master equation in \cite{blencowe} based on a scalar field, incorporates both vacuum and thermal gravitational fluctuations and can in principle be applied to relativistic domains as with Eq. \eqref{maseqn}. However, the gravitational fields considered there are gauge fixed and account only for graviton effects. Our approach encompasses the full decomposition of gravity including consistently self-gravity associated with general bosonic fields. It provides a general theory beyond other influence functional models such as \cite{garay, breuer1} based on simplified stochastic noise models with Markovian assumptions.

The versatility of Eq. \eqref{maseqn} allows one to probe a range of gravitational fluctuation scenarios including those due to vacuum fluctuations, subject to a cutoff $\Omega$ discussed above, as suggested by a number of authors including \cite{blencowe, anastopoulos}. As an illustrative example, let us consider light from an incoherent source like a star, emitting photons up to a maximum frequency $\omega_*$, which will be regarded as the maximum cutoff. Though \eqref{maseqn} is amenable to indefinite particle numbers,
for simplicity, let us treat the light system to consist of one-photon states $\ket{\kk,\lambda}$ with $\lambda=1,2$ and wave-vector $\kk$ for $k\le k_*$ relative to the light source.
The matrix elements of the normal-ordered TT Maxwell stress tensor are calculated to be
\begin{eqnarray*}
&&
\bra{\kk_1,\lambda} \tau_{ij}(\kk,t) \ket{\kk_2,\sigma}
=
-\hbar\,
\delta^3(\kk_2-\kk_1-\kk)
\nppp&&
\hspace{35pt}
\times\; P_{ijkl}(\kk)
\sqrt{{k_1}{k_2}}\,
E^{\lambda\sigma}_{kl}(\kk_1,\kk_2)
e^{i({k_1}-{k_2})t}
\end{eqnarray*}
where
$E^{\lambda\sigma}_{ij}(\kk,\kk')
=
e_{i}^{\lambda}(\kk) e_{j}^{\sigma}(\kk')
+
\epsilon_{\lambda\lambda'}\epsilon_{\sigma\sigma'}
e_{i}^{\lambda'}(\kk) e_{j}^{\sigma'}(\kk')
$
using the polarization vectors $e_{i}^{\lambda}(\kk)$
normalized with $e_{i}^\lambda(\kk)e_{j}^\lambda(\kk)=P_{ij}(\kk)$.
In this simple example, let us neglect self-gravity and take zero temperature.
Using photon states with incoherent polarizations so that
$\bra{\kk,\lambda}\rho(t)\ket{\kk,\sigma}=\rho(\kk,t)\delta_{\lambda\sigma}/2$, we obtain the decoupled diagonal part of \eqref{maseqn} as follows
\begin{eqnarray}
&&\hspace{-20pt}
\dot\rho(\kk,t)
=
-8\pi G\hbar
\int
\frac{\dd^3k'\,{k k'}H(k_*-k')}{2(2\pi)^3|\kk'-\kk|} P_{ijkl}(\kk'-\kk)
\nppp&&\!\!\!\!\!\!\!\!\!\times\,
(E^{\lambda\sigma}_{ij}E^{\lambda\sigma}_{kl})(\kk,\kk')\!
\left[
\frac{\sin\chi t}{\chi}\rho(\kk,t)
-
\frac{\sin\chi't}{\chi'}\rho(\kk',t)
\right]
\label{prpg}
\end{eqnarray}
where $\chi=|\kk'-\kk|+{k'}-{k}$, $\chi'=|\kk'-\kk|-{k'}+{k}$.
The conservation of total probability, $\int\dd^3k\,H(k_*-k)\rho(\kk,t)=1$, is clearly preserved owing to the symmetry between $\kk$ and $\kk'$ in the integrand of \eqref{prpg}.

It is instructive to observe how the initial profile of $\rho(\kk,t)$ affects its dissipation rate over the initial period of $0 < t < 1/k_*$. First we note that for a flat spectrum with constant $\rho(\kk,0)$ we see immediately $(\dot\rho/\rho)(\kk,0)\approx 0$ which is to be expected as it represents a completely dissipated state in vacuum at zero temperature.
We then consider a narrow band spectrum that peaks around the maximum momentum $k_*$. As an order-of-magnitude estimate, we find after evaluating \eqref{prpg} that the corresponding initial dissipation rate with the speed of light $c$ restored is given by
\begin{eqnarray}
-(\dot\rho/\rho)(\kk_*,0)
\lesssim
t_P^2\, \omega_*^3
\label{litdrate}
\end{eqnarray}
in terms of the Planck time $t_P$ and photon frequency $\omega_*=ck_*$. A similar estimate for a massive scalar satisfying the dispersion relation $\omega^2= c^2(k^2+\mu^2)$ with a mass parameter $\mu\ge0$, yields
\begin{eqnarray}
-(\dot\rho/\rho)(\kk_*,0)
\lesssim
\frac{c^3 t_P^2\, k_*^4}{\sqrt{k_*^2+\mu^2}}
\label{scldrate}
\end{eqnarray}
showing a quartic momentum dependence of the initial quantum gravitational dissipation for a nonrelativistic particle tending to a cubic dependence as in \eqref{litdrate}, when it becomes relativistic or indeed massless.

As a relatively simple application of the master equation \eqref{maseqn}, the potential size of effects in \eqref{litdrate} and \eqref{scldrate} is unsurprisingly small on typical laboratory scales.
Since the possible dressing of states due to gravitational quantum vacuum could partially or fully suppress vacuum dissipation, inequalities \eqref{litdrate} and \eqref{scldrate} serve as a possible upper bound that may guide future work.
Nonetheless, their origins in quantum gravity are significant.
Similar quantum effects are known to be collectively amplified with a large number of correlated and identical particles \cite{breuer2}, which may lead to observational implications of the quantum gravitational dissipation of starlight.
Further development of this work may be relevant for
addressing
the emergence of the classical behaviour of spacetime and the origin of the initial perturbations in cosmology \cite{kiefer2} and in relativistic astrophysics \cite{landulfo}.

\section*{ACKNOWLEDGMENTS}

T.O. is most grateful to the Carnegie Trust for the Universities of Scotland for a scholarship. C.W. acknowledges support from the EPSRC GG-Top Project.

\end{document}